\documentclass[pdflatex,sn-mathphys]{sn-jnl}

\jyear{2021}%

\usepackage{graphicx}
\usepackage{dcolumn}
\usepackage{bm}
\usepackage{hyperref}

\usepackage{amsmath,amssymb,amsthm,easybmat,verbatim}
\usepackage{enumerate}
\usepackage{hyperref}
\usepackage{color}
\usepackage{graphicx}
\usepackage{array}
\usepackage{float}

\usepackage{physics}
\usepackage{mathtools}

\theoremstyle{thmstyleone}%
\theoremstyle{thmstyletwo}%

\theoremstyle{thmstylethree}%

\begin{document}

\title{Identifying Quantum Correlations Using Explicit SO(3) to SU(2) Maps}

\author*[1,4]{\fnm{Daniel} \sur{Dilley}}\email{quantumdilley@yahoo.com}

\author[3]{\fnm{Alvin} \sur{Gonzales}}

\author[1,2]{\fnm{Mark} \sur{Byrd}}

\affil*[1]{\orgdiv{School of Physics and Applied Physics}, \orgname{Southern Illinois University Carbondale}, \orgaddress{\street{1245 Lincoln Drive}, \city{Carbondale}, \postcode{62901}, \state{IL}, \country{USA}}}

\affil[2]{\orgdiv{School of Computing}, \orgname{Southern Illinois University Carbondale}, \orgaddress{\street{1230 Lincoln Drive}, \city{Carbondale}, \postcode{62901}, \state{IL}, \country{USA}}}

\affil[3]{\orgdiv{Intelligence Community Postdoctoral Research Fellowship Program}, \orgname{Argonne National Laboratory}, \orgaddress{\street{9700 S. Cass Avenue}, \city{Lemont}, \postcode{60439}, \state{IL}, \country{USA}}}

\affil[4]{\orgdiv{Mathematics Department}, \orgname{University of Arkansas-Fort Smith}, \orgaddress{\street{5210 Grand Avenue}, \city{Fort Smith}, \postcode{72913}, \state{AR}, \country{USA}}}



\abstract{
Quantum state manipulation of two-qubits on the local systems by special unitaries induces special orthogonal rotations on the Bloch spheres. An exact formula is given for determining the local unitaries for some given rotation on the Bloch sphere. The solution allows for easy manipulation of two-qubit quantum states with a single definition that is programmable. With this explicit formula, modifications to the correlation matrix are made simple. Using our solution, it is possible to diagonalize the correlation matrix without solving for the parameters in SU(2) that define the local unitary that induces the special orthogonal rotation in SO(3). Since diagonalization of the correlation matrix is equivalent to diagonalization of the interaction Hamiltonian, manipulating the correlation matrix is important in time-optimal control of a two-qubit state. The relationship between orthogonality conditions on SU(2) and SO(3) are given and manipulating the correlation matrix when only one qubit can be accessed is discussed. 
}




\maketitle

\section{Introduction}

A qubit is the fundamental component of quantum information and its properties allow us to perform tasks that classical machines are incapable of executing. One strength of a qubit is tied to the quantum gates that can effectively create superpositions and thus create entangled states. Entangling qubits gives quantum devices the capability of teleportation \cite{Bennet_1993}, super-dense coding \cite{Bennet_1992}, unstructured search \cite{Grover_1996}, prime factorization \cite{Shor_1994}, and the ability to perform many other quantum algorithms and protocols that are not possible classically. The ability to use nonlocal correlations in quantum protocols allow advantages over their classical counter parts \cite{Parakh_2022,cerf_2000,Gottesman_1999}. 
Given the various applications of quantum information processing, it is crucial that we understand the quantum correlations and the ways we can manipulate them.  

It is conventional to apply local unitaries to a two-qubit system without regard to how the final correlations between the systems will appear. This paper aims to take a different viewpoint and show how the correlations can be manipulated to have a particular form using local unitary transformations applied to the sub-systems.  If there is knowledge about how two single-qubit systems are correlated \cite{Blume_Kohout_2010,Li_2017}, then ideally one could change those correlations to make them symmetric or eliminate some of the elements of the correlation matrix.  

It turns out that the correlations between two single-qubit systems is all that is needed in some information tasks such as Bell inequality violations \cite{Horodecki_1995} or witnessing entanglement using projective measurements \cite{Hyllus_2005}. By a correct choice of measurements, in two distant labs that share an entangled two-qubit state, nonlocality can be demonstrated. This is equivalent to rotating the local Bloch vectors first, hence rotating the correlation matrix, and then making a simple Pauli-Z measurement. By having the ability to rotate the local states, one can always measure along any axis they choose. This is demonstrated by realizing that local orthogonal rotations on the correlation matrix $\mathcal{T}_\rho$ do not change the eigenvalues of the matrix $\mathcal{T}_\rho \cdot \mathcal{T}_\rho^T$ given in \cite{Horodecki_1995}. Since the optimal violation of the Bell CHSH inequality only depends on the square of the sum of the two largest eigenvalues of this symmetric matrix, it does not change with local unitaries.

In addition to the state of a quantum system, the Hamiltonian of two two-state systems has the same mathematical form as the density operator, excluding the constraints of positivity and trace one.  When systems interact with each other they can become entangled and some entangling gates are better at creating entanglement than others \cite{Zhang_2013}. Furthermore, some gates are time-optimal when considering their ability to produce correlations when the single particle operations are much faster than the interaction Hamiltonian  \cite{Khaneja_2001}. For example, this happens in spin systems. In addition, such considerations are important in protocols where local unitary transformations are available and non-local ones are not. Thus, the non-local part of the operator can be the most important part whether it is a correlation matrix of a density operator or an interaction Hamiltonian.  

In this paper we provide explicit formulas for diagonalizing the correlation matrix of the density operator, or equivalently, diagonalizing the interaction Hamiltonian.  This is done by explicit formulas for the SU(2) rotation matrices given the SO(3) rotation matrices in the adjoint representation of SU(2).  In Section II, we review some material regarding the transformation of SU(2) to SO(3) and also show the relation between unit quaternions and $2\times2$ complex matrices in SU(2). It will become evident in Section III why the scalar and real parts of a quaternion are important when considering the transformation from SO(3) to SU(2). The explicit formula is given with additional work proofs provided in the Appendix. In Section IV, we illustrate the power of having an explicit transformation from SO(3) to SU(2) by using our formula to diagonalize the correlation matrix in our example. The conditions for orthogonality between two elements in SU(2) or two elements in SO(3) are given in Section V and then we move on to Section VI to discuss the constraint of only having access to one qubit of a two-qubit state as given in Eq. {\ref{Eq:Two_Qubit_State}}. In Section VII, we provide a link to a Mathematica program that includes our explicit formula for anyone who would like to download and use for their own purposes. Lastly, in Section \ref{sec:conclusion} we summarize our results.

\section{Background}

A general two-qubit state $\rho^{AB}$ can be defined in terms of a $3 \times 3$ correlation matrix with elements $\{ \mathcal{T}_\rho \}_{ij} = t_{ij} = \tr(\hat{\sigma}_i \otimes \hat{\sigma}_j \cdot \rho^{AB})$ and two local Bloch vectors $\vec{a}$, $\vec{b}$ that define the states $\rho^A$, $\rho^B$ of systems $A,B$. These vectors are defined as $a_i = \tr(\hat{\sigma}_i \otimes \mathbb{I} \cdot \rho^{AB})$ and $b_j = \tr(\mathbb{I} \otimes \hat{\sigma}_j \cdot \rho^{AB})$ and the overall state is expressed as
\begin{align} \label{Eq:Two_Qubit_State}
    \rho^{AB} = \dfrac{1}{4} \left( \mathbb{I} \otimes \mathbb{I} + \vec{a} \cdot \vec{\sigma} \otimes \mathbb{I} + \mathbb{I} \otimes \vec{b} \cdot \vec{\sigma} + \sum_{i,j = 1}^3 t_{ij} \hat{\sigma}_i \otimes \hat{\sigma}_j \right)
\end{align}
for the vector of Pauli matrices $\vec{\sigma} = (\hat{\sigma}_1,\hat{\sigma}_2,\hat{\sigma}_3)$. A local unitary on sub-system $A$ will induce a three-dimensional rotation to $\vec{a}$ and the left side of the correlation matrix $\mathcal{T}_\rho$ will get hit by the same transformation. On the other hand, a local unitary acting on sub-system $B$ will induce a three-dimensional rotation to the $\vec{b}$, but the the transpose of that transformation will act on the right side of the correlation matrix $\mathcal{T}_\rho$.

If two systems, $A$ and $B$ interact, then the Hamiltonian that governs their evolution is given by the equation
\begin{align} \label{Eq:Hamiltonian}
    H^{AB} = \mathbb{I}^A \otimes \mathbb{I}^B + H^A \otimes \mathbb{I}^B + \mathbb{I}^A \otimes H^B + \sum_{i,j} R_i^A \otimes S_j^B
\end{align}
for the operators $\{H^A,R_i^A\}$ that solely act on system $A$ and the operators $\{ H^B, S_j^B \}$ that solely act on system $B$. The term $\sum_{i,j} R_i^A \otimes S_j^B$ is the interaction Hamiltonian that couples the two subsystems.  By exponentiation of $H^{AB}$, we see that the identity part of Eq. (\ref{Eq:Hamiltonian}) only introduces a global phase. In reference \cite{Khaneja_2001}, it was shown that diagonalizing the interaction Hamiltonian leads to conditions for the time-optimal control of a state of two-qubits. This is equivalent to diagonalizing the correlation matrix in Eq. (\ref{Eq:Two_Qubit_State}).

For the reverse direction of going from SU(2) to SO(3), it is well known that we can use the transformation (see for example \cite{cornwell1984})
\begin{align} \label{Eq:su2toso3}
    \mathcal{O}_{ij} = \dfrac{1}{2} \text{tr}[\hat{\sigma}_i U \hat{\sigma}_j U^\dagger],
\end{align}
where $\mathcal O \in \text{SO(3)}$ and $U \in \text{SU(2)}$, such that $\mathcal{O}(U) = \mathcal{O}(-U)$ due to the double cover of SU(2) to SO(3). For the right unitary, we would simply have $\mathcal{O}_{ji}$ since it induces the transpose of $\mathcal{O}$ on the right side of $\mathcal{T}_\rho$. Specifically, the transformations take the form $\vec{a} \rightarrow L \cdot \vec{a}$, $\vec{b} \rightarrow R \cdot \vec{b}$, and $\mathcal{T}_\rho \rightarrow L \cdot \mathcal{T}_\rho \cdot R^T$ from the local unitaries $U_L \otimes U_R \in \text{SU}(2) \times \text{SU}(2)$ applied to $\rho^{AB}$ \cite{Makhlin_2002}. Thus, after diagonalizing the correlation matrix $\mathcal{T}_\rho$ with our unitaries $U_L$ and $U_R$, we can easily find the special orthogonal rotations $\mathcal{O}_L$ and $\mathcal{O}_R$ from Eq. (\ref{Eq:su2toso3}) to see how the local Bloch vectors rotate as well.

There is a nice representation of complex matrices in SU(2) in terms of quaternions \cite{Hamilton_1843} which can be spanned by the matrices $\{ \mathbb{I}, i\; \hat{\sigma}_1, i \; \hat{\sigma}_2,i\; \hat{\sigma}_3 \} = \{\bf{1},\bf{i},\bf{j},\bf{k}\}$ which have the properties $\bf{i}^2 = \bf{j}^2 = \bf{k}^2 = \bf{ijk} = - \bf{1}$ and 
\begin{align}
    \bf{ij} = \bf{k} = -\bf{ji} \quad \bf{jk} = \bf{i} = -\bf{kj} \quad \bf{ki} = \bf{j} = -\bf{ik}.
\end{align}
Note that the combination
\begin{align}
    q = \alpha_1 + \alpha_2 \bf{i} + \beta_1 \bf{j} + \beta_2 \bf{k} = 
    \left(
    \begin{array}{cc}
        \alpha_1 + i \; \alpha_2 & \beta_1 + i \; \beta_2 \\
        -\beta_1 + i \; \beta_2 & \alpha_1 - i \; \alpha_2
    \end{array}
    \right)
\end{align}
gives an arbitrary element of SU(2) for when the quaternion has norm 1; that is
\begin{align}
    \sqrt{q^\dagger q} = 1 \Rightarrow \alpha_1^2 + \alpha_2^2 + \beta_1^2 + \beta_2^2 = 1.
\end{align}
We say that the $\it{scalar}$ or $\it{real}$ part of a quaternion is given by $\alpha_1$ and that the $\it{vector}$ or $\it{imaginary}$ part is given by $\alpha_2 \bf{i} + \beta_1 \bf{j} + \beta_2 \bf{k}$.  In the case that $\alpha_1=0$, the solution is quite straight-forward.  However, in the case that $\alpha_1\neq 0$, several cases must be considered separately.  These are specified in the next section.




\section{Explicit construction of SU(2) from SO(3)}
Let us assume that the initial local unitary acting on a general two-qubit state is given by an arbitrary SU(2)
\begin{align} \label{Eq:Local_Unitary_in_SO(2)}
    U = \pm
    \left(
    \begin{array}{cc}
        \alpha_1 + i \; \alpha_2 & \beta_1 + i \; \beta_2  \\
        -\beta_1 +i \; \beta_2 & \alpha_1 - i \; \alpha_2 
    \end{array}
    \right)
\end{align}
so that the corresponding matrix in SO(3) is given by the Euler–Rodrigues formula \cite{Euler_1771}
\begin{align}
    &\mathcal{O} = \left(
        \begin{array}{ccc}
            \tau_1 & 2(\mu_{12} + \nu_{12}) & 2(-\chi_{11} + \chi_{22}) \\
            2(-\mu_{12} + \nu_{12}) & \tau_2 & 2(\chi_{21} + \chi_{12}) \\
            2(\chi_{11} + \chi_{22}) & 2(\chi_{21} - \chi_{12}) & \tau_3
        \end{array}
    \right)
\end{align}
for $\chi_{ij} = \alpha_i \beta_j$, $\mu_{ij} = \alpha_i \alpha_j$, $\nu_{ij} =\beta_i \beta_j$, and
\begin{align}
    \tau_1 &= \alpha_1^2 - \alpha_2^2 - \beta_1^2 + \beta_2^2, \;
    \tau_2 = \alpha_1^2 - \alpha_2^2 + \beta_1^2 - \beta_2^2, 
    \tau_3 = \alpha_1^2 + \alpha_2^2 - \beta_1^2 - \beta_2^2
\end{align}
according to Eq. (\ref{Eq:su2toso3}). This operator describes an arbitrary rotation of a three-dimensional vector given by $w' = \mathcal{O} \cdot w$. Now we define the maximally entangled Bell states to be
\begin{align} \nonumber
    &\ket{\Phi^+} = \dfrac{1}{\sqrt{2}} \left( \ket{00} + \ket{11} \right) \qquad \ket{\Phi^-} = \dfrac{1}{\sqrt{2}} \left( \ket{00} - \ket{11} \right) \\ \nonumber
    &\ket{\Psi^+} = \dfrac{1}{\sqrt{2}} \left( \ket{01} + \ket{10} \right) \qquad \ket{\Psi^-} = \dfrac{1}{\sqrt{2}} \left( \ket{01} - \ket{10} \right)
\end{align}
and we use the basis 
\begin{align}
    L_1 = 
    \left(
    \begin{array}{ccc}
        0 & -1 & 0 \\
        1 & 0 & 0 \\
        0 & 0 & 0
    \end{array}
    \right)&, 
    L_2 = 
    \left(
    \begin{array}{ccc}
        0 & 0 & 1 \\
        0 & 0 & 0 \\
        -1 & 0 & 0
    \end{array}
    \right),
    \text{and } L_3 = 
    \left(
    \begin{array}{ccc}
        0 & 0 & 0 \\
        0 & 0 & -1 \\
        0 & 1 & 0
    \end{array}
    \right)
\end{align}
for the Lie algebra of $\mathfrak{so}(3)$ \cite{Hall_2015}. Define the following sign function:
\begin{align}
    \text{sgn}(t) = \begin{cases}
                        -1 &\text{ if } t < 1 \\
                        0  &\text{ if } t = 0 \\
                        1  &\text{ if } t > 1
                    \end{cases}
\end{align}
Then notice that
\begin{align}
    \text{sgn}[\tr (\mathcal{O} \cdot L_1)] &= \text{sgn}[\alpha_1 \alpha_2] = \text{sgn}[\alpha_1] \cdot \text{sgn}[\alpha_2] \\
    \text{sgn}[\tr (\mathcal{O} \cdot L_2)] &= \text{sgn}[\alpha_1 \beta_1] = \text{sgn}[\alpha_1] \cdot \text{sgn}[\beta_1] \\
    \text{sgn}[\tr (\mathcal{O} \cdot L_3)] &= \text{sgn}[\alpha_1 \beta_2] = \text{sgn}[\alpha_1] \cdot \text{sgn}[\beta_2] \\
    \text{sgn}[\tr (\mathcal{O} \cdot \abs{L_1})] &= \text{sgn}[\beta_1 \beta_2] = \text{sgn}[\beta_1] \cdot \text{sgn}[\beta_2] \\
    \text{sgn}[\tr (\mathcal{O} \cdot \abs{L_2})] &= \text{sgn}[\alpha_2 \beta_2] = \text{sgn}[\alpha_2] \cdot \text{sgn}[\beta_2] \\
    \text{sgn}[\tr (\mathcal{O} \cdot \abs{L_3})] &= \text{sgn}[\alpha_2 \beta_1] = \text{sgn}[\alpha_2] \cdot \text{sgn}[\beta_1]
\end{align}
and 
\begin{equation}\label{eq:tracecond}
1 + \tr (\mathcal{O}) = 4\alpha_1^2
\end{equation}
which implies that if $\alpha_1 = 0$, then $\tr (\mathcal{O}) = -1$. This means that the corresponding $q$ must be a vector quaternion if $\tr (\mathcal{O}) = -1$. We then make the following calculations
\begin{align} \nonumber
    \dfrac{1}{2} \sqrt{1 - \tr (\mathcal{O} \cdot \mathcal{T}_{\Psi^-})} &= \sqrt{\alpha_1^2}, \; \dfrac{1}{2} \sqrt{1 - \tr (\mathcal{O} \cdot \mathcal{T}_{\Psi^+})} = \sqrt{\alpha_2^2} \\
   \dfrac{1}{2} \sqrt{1 - \tr (\mathcal{O} \cdot \mathcal{T}_{\Phi^+})} &= \sqrt{\beta_1^2}, \; \dfrac{1}{2} \sqrt{1 - \tr (\mathcal{O} \cdot \mathcal{T}_{\Phi^-})} = \sqrt{\beta_2^2}
\end{align}
where $\mathcal{T}_{\rho}$ is the correlation matrix of the state $\rho$ (see Eqs. \eqref{eq:corr_mats1} and \eqref{eq:corr_mats2} in the Appendix for the matrices). Now we can put all of these results together to get the exact closed-form solution for a pair of local special unitaries $\{U,-U\}$ that will induced a special orthogonal matrix $\mathcal{O} \in$ SO(3) if $\tr (\mathcal{O}) \neq -1$.

By performing the above operations, we get the matrix $\text{sgn}[\alpha_1] \cdot (\pm U) = \pm U$ since $\text{sgn}[\kappa] \cdot \sqrt{\kappa^2} = \kappa$ for any $\kappa$ and $\text{sgn}[\alpha_1] = \pm 1$. The reason our solution does not work for vector quaternions is that if $\text{sgn}[\alpha_1] = 0$, then we get the zero matrix. Thus, if an element $\mathcal{O} \in SO(3)$ is associated with a quaternion that contains a real part, then the exact solutions are given by
\begin{align}
    \label{Eq:U_Sol_General}
    \pm U_R (\mathcal{O}) = 
    \pm \text{sgn}(\alpha_1) \left(
    \begin{array}{cc}
        \alpha_1 (\mathcal{O}) + i \; \alpha_2 (\mathcal{O}) & \beta_1 (\mathcal{O}) + i \; \beta_2 (\mathcal{O}) \\
        -\beta_1 (\mathcal{O}) + i \; \beta_2 (\mathcal{O}) & \alpha_1 (\mathcal{O}) - i \; \alpha_2 (\mathcal{O})
    \end{array}
    \right) \in SU(2)
\end{align}
for
\begin{align}
    \abs{\alpha_1 (\mathcal{O})} &= \dfrac{1}{2} \sqrt{1 - \tr (\mathcal{O} \cdot \mathcal{T}_{\Psi^-})} \\
    \text{sgn}(\alpha_1) \cdot \alpha_2 (\mathcal{O}) &= \dfrac{1}{2} \text{sgn}[\tr (\mathcal{O} \cdot L_1)] \sqrt{1 - \tr (\mathcal{O} \cdot \mathcal{T}_{\Psi^+})} \\
    \text{sgn}(\alpha_1) \cdot \beta_1 (\mathcal{O}) &= \dfrac{1}{2} \text{sgn}[\tr (\mathcal{O} \cdot L_2)] \sqrt{1 - \tr (\mathcal{O} \cdot \mathcal{T}_{\Phi^+})} \\
    \text{sgn}(\alpha_1) \cdot \beta_2 (\mathcal{O}) &= \dfrac{1}{2} \text{sgn}[\tr (\mathcal{O} \cdot L_3)] \sqrt{1 - \tr (\mathcal{O} \cdot \mathcal{T}_{\Phi^-})}.
\end{align}

Now what would the solution be if the quaternion had no real part; that is, how would the formula change if $\tr (\mathcal{O}) = -1$? A general solution would then be given by
\begin{align}
    U(\mathcal{O}) = \pm \left(U_R (\mathcal{O}) +  (1-\text{sgn}[1+\tr (\mathcal{O})]) \cdot U_V(\mathcal{O}) \right)
\end{align}
for the vector part $U_V$ when $\tr (\mathcal{O}) = -1$. This will give an exact closed-form solution for determining the pair of local unitaries in SU(2) that will induce the orthogonal matrix $\mathcal{O} \in$ SO(3) when it acts on one of the local systems of a two-qubit state $\rho^{AB}$. To determine $U_V$ for when certain parameters can be equal to zero, we define the matrix function
\begin{align}
    W&(\mathcal{O},x,y,z) = 
    \left(
    \begin{array}{cc}
        i \; a_2(\mathcal{O},x) & b_1(\mathcal{O},y) + i \; b_2(\mathcal{O},z) \\
        -b_1(\mathcal{O},y) + i \; b_2(\mathcal{O},z) & - i \; \alpha_2(\mathcal{O},x)
    \end{array}
    \right)
\end{align}
for the values
\begin{align}
    a_2(\mathcal{O},x) &= \dfrac{1}{2} \text{sgn}[\tr (\mathcal{O} \cdot x)] \sqrt{1 - \tr (\mathcal{O} \cdot \mathcal{T}_{\psi^+})} \\
    b_1(\mathcal{O},y) &= \dfrac{1}{2} \text{sgn}[\tr (\mathcal{O} \cdot y)] \sqrt{1 - \tr (\mathcal{O} \cdot \mathcal{T}_{\Phi^+})} \\
    b_2(\mathcal{O},z) &= \dfrac{1}{2} \text{sgn}[\tr (\mathcal{O} \cdot z)] \sqrt{1 - \tr (\mathcal{O} \cdot \mathcal{T}_{\Phi^-})}
\end{align}
and let $\abs{A}$ be the absolute value matrix with elements $\abs{A_{ij}}$. 

The general solution of $U_V$ is proven in Appendix \ref{App:V_Sol_Proof} and is given by
\begin{align} \nonumber
    U_V(\mathcal{O}) =& \; W(\mathcal{O},\abs{L_1},\abs{L_2},\abs{L_3}) \nonumber + \; (1-\gamma_1) \gamma_2 \gamma_3 \cdot W(\mathcal{O},\mathbb{I},\mathbb{I},-\abs{L_1}) \nonumber \\ 
    +& \; \gamma_1 (1-\gamma_2) \gamma_3 \cdot W(\mathcal{O},-\abs{L_2},\mathbb{I},\mathbb{I}) \nonumber
    + \; \gamma_1 \gamma_2 (1-\gamma_3) \cdot W(\mathcal{O},\mathbb{I},-\abs{L_3},\mathbb{I}) \nonumber \\ \label{Eq:V_Sol}
    +& \; \gamma_1 \gamma_2 \gamma_3 \cdot W(\mathcal{O},-\mathbb{I},-\mathbb{I},-\mathbb{I})
\end{align}
for 
\begin{align}
    \gamma_i = 1 - \text{sgn}[\tr (\mathcal{O} \cdot \abs{L_i})]^2.
\end{align}
Therefore, the general solution is given by
\begin{align} \label{Eq:General_Solution}
    U(\mathcal{O}) = \pm \left( U_R(\mathcal{O}) + (1-\text{sgn}[1+\tr (\mathcal{O})]^2) \cdot U_V(\mathcal{O}) \right)
\end{align}
where $U_R(\mathcal{O})$ is our solution given in Eq. (\ref{Eq:U_Sol_General}), which in most realistic scenarios is just equal to $U_R(\mathcal{O})$ since there is a very high probability that $\tr (\mathcal{O}) \neq -1$ for a two-qubit state prepared in a lab. Some small perturbation error would almost certainly cause the real part of the quaternion, associated with the local unitary, to be non-zero. The gamma functions simply pull out each special case to avoid incorrect solutions from repetitions. This solution is simple to program and can save much time when calculating a local unitary in SU(2) that induces some wanted orthogonal rotation in SO(3) that rotates the Bloch vector and correlation matrix of a two-qubit quantum state. 

In summary, the general solution is given by
\begin{align}
    U(\mathcal{O}) = \begin{cases}
                        \pm U_R (\mathcal{O}) \text{ if } \tr (\mathcal{O}) \neq -1 \\
                        \pm U_V (\mathcal{O}) \text{ if } \tr (\mathcal{O}) = -1
                     \end{cases}
\end{align}
which can be written compactly as Eq. (\ref{Eq:General_Solution}). We see that if $\alpha_1 = 0$ then $\tr (\mathcal{O}) = -1$. This means that $U_R(\mathcal{O})$ would be zero according to Eq. (\ref{Eq:U_Sol_General}) and that the coefficient $(1-\text{sgn}[1+\tr (\mathcal{O})]^2) = 1$. Therefore, the solution in Eq. (\ref{Eq:General_Solution}) either takes on the form of $U_R (\mathcal{O})$ or $U_V (\mathcal{O})$. The matrix $U_R (\mathcal{O})$ is always the correct solution whenever the real part of the quaternion is non-zero and the matrix $U_V (\mathcal{O})$ is always the solution whenever the real part of the quaternion is zero. Our general explicit formula ensures both cases are mutually exclusive. 

We provide a program in Mathematica that we use to define the general solution and go over a simple example for how to diagonalize the correlation matrix. Access to the repository is provided in Section \ref{sec:DataAvail}. 
We will give some examples in the next section on how we can use this formula.


\section{Diagonalizing the Correlation Matrix}

If we want to diagonalize any correlation matrix of $\rho$, we first use the singular value decomposition (SVD) of $\mathcal{T}_\rho$.  To rewrite it with $SO(3)$ on the outsides of the decomposition, we modify the SVD and put the decomposition in the form $L\Sigma R$, where $L,R\in SO(3)$ and $\Sigma$ is a diagonal matrix with not necessarily positive entries. 
We must then apply the local rotations $U_L \otimes U_R$ to our state so that the correlation matrix becomes $L^T L \Sigma R R^T = \Sigma$. Keep in mind that the local unitary on the second sub-system induces a right special orthogonal $3 \times 3$ to the correlation matrix that is transposed. This solution is not only good for any general two-qubit state $\rho$, but it is also readily adapted to almost any programming language.  

Suppose we want to locally rotate the initial entangled state
\begin{align}
\rho^{AB} = 
\dfrac{1}{4}
    \left(
    \begin{array}{cccc}
        1 & -1 & i & i \\
        -1 & 1 & -i & -i \\
        -i & i & 1 & 1 \\
        -i & i & 1 & 1
    \end{array}
    \right)
\end{align}
so that its correlation matrix
\begin{align}
    L \Sigma R = 
    \left(
    \begin{array}{ccc}
        0 & 0 & 1 \\
        1 & 0 & 0 \\
        0 & 1 & 0  
    \end{array}
    \right)
    \cdot
    \left(
    \begin{array}{ccc}
        1 & 0 & 0 \\
        0 & -1 & 0 \\
        0 & 0 & 1 
    \end{array}
    \right)
    \cdot
    \left(
    \begin{array}{ccc}
        0 & 0 & -1 \\
        1 & 0 & 0 \\
        0 & -1 & 0
    \end{array}
    \right)
\end{align}
is diagonal. Using Eq. (\ref{Eq:General_Solution}) for both $L^T$ and $R$, we get the special unitaries
\begin{align}
    U_{L^T} = 
    \dfrac{1}{2}
    \left(
    \begin{array}{cc}
        1+i & 1+i \\
        -1+i & 1-i
    \end{array}
    \right), \;
    U_R = 
    \dfrac{1}{2}
    \left(
    \begin{array}{cc}
        1-i & 1+i \\
        -1+i & 1+i
    \end{array}
    \right) 
\end{align}
that induce the transformation $L^T (L \Sigma R) R^T \rightarrow \Sigma$ and we are left with the maximally entangled Bell state $\Phi^+$ since the local Bloch vectors were initially equal to $\vec{0}$.

What if we wanted to rotate the local Bloch vector of system A about the $x,y,$ or $z$-axis at some angle $\theta$? The rotations in $SO(3)$ have the form
\begin{align} \nonumber
    \mathcal{X} = 
    \left(
    \begin{array}{ccc}
        1 & 0 & 0 \\
        0 & \text{cos}(\theta) & -\text{sin}(\theta) \\
        0 & \text{sin}(\theta) & \text{cos}(\theta)
    \end{array}
    \right)&,
    \mathcal{Y} = 
    \left(
    \begin{array}{ccc}
        \text{cos}(\theta) & 0 & \text{sin}(\theta) \\
        0 & 1 & 0 \\
        -\text{sin}(\theta) & 0 & \text{cos}(\theta)
    \end{array}
    \right), \\
    \text{and } \mathcal{Z} = 
    &\left(
    \begin{array}{ccc}
        \text{cos}(\theta) & -\text{sin}(\theta) & 0 \\
        \text{sin}(\theta) & \text{cos}(\theta) & 0 \\
        0 & 0 & 1
    \end{array}
    \right)
\end{align}
which have an associated $SU(2)$ representation of
\begin{align} \nonumber
    x_A = \pm
    \left(
    \begin{array}{cc}
        \text{cos}(\theta/2) & -i \; \text{sin}(\theta/2) \\
        -i \; \text{sin}(\theta/2) &  \text{cos}(\theta/2)
    \end{array}
    \right),
    &y_A = \pm
    \left(
    \begin{array}{cc}
        \text{cos}(\theta/2) & -\text{sin}(\theta/2) \\
        \text{sin}(\theta/2) & \text{cos}(\theta/2)
    \end{array}
    \right), \\
    \text{and } z_A = \pm&
    \left(
    \begin{array}{cc}
        e^{-i \theta/2} & 0 \\
        0 & e^{i \theta/2}
    \end{array}
    \right)
\end{align}
given by the (+) solution of Eq. \eqref{Eq:General_Solution}. 
If $\theta \in [0,\pi)$, then the solution is the set of matrices with plus signs.  If $\theta \in [\pi,2\pi)$, then the solution is the set of matrices with the minus signs. Either rotation will transform the local Bloch vector appropriately since $SU(2)$ double covers $SO(3)$, as seen in Eq.~(\ref{Eq:su2toso3}) when switching $U$ with $-U$. Since all rotations are explained in terms of these rotations, it is easy to verify Eq.~(\ref{Eq:General_Solution}).


\section{Orthogonality}

Let $U_1,U_2\in SU(2)$ and $\mathcal{O}_1, \mathcal{O}_2 \in \text{Adj}(SU(2))\cong  SO(3)$, the adjoint representation of $SU(2)$. Then if $\tr(U_1 U_2^\dagger) = 0$, we say these are orthogonal matrices.  One may ask, what is the condition for the corresponding $\mathcal{O}_1$ and $\mathcal{O}_2$ matrices? One way to find the condition is to rely on representation-theoretic argument as in \cite{Byrd_2011}.  The argument is as follows. The tensor product of two $U\in SU(2)$ is a reducible representation and can be reduced to a three dimensional and one dimensional representation.  Then, the following shows the condition: 
\begin{eqnarray}
(U_1\otimes U_1) (U_2^\dagger\otimes U_2^\dagger) &=& (U_1U_2^\dagger)\otimes (U_1U_2^\dagger) \\
&=& 
(\mathcal{O}_1 \oplus 1)(\mathcal{O}_2^T\oplus 1) \\
&=& (\mathcal{O}_1 \mathcal{O}_2^T \oplus 1),
\end{eqnarray}
where the second line follows from the decomposition of the tensor product.   
Taking the trace of the first and last expressions, and given that  $\mbox{Tr}(U_1U_2^\dagger) = 0$, we get $\mbox{Tr}(\mathcal{O}_1 \mathcal{O}_2^T)=-1$.  

Another way to show this is quite straight-forward given the results above.  Examining  Eq.~(\ref{eq:tracecond}): 
\begin{equation}
1 + \tr (\mathcal{O}) = 4\alpha_1^2
\end{equation}
we see that $\alpha_1 = 0$ implies $\tr (\mathcal{O}) = -1$. As can be seen from  Eq.~(\ref{Eq:Local_Unitary_in_SO(2)}), the trace of the unitary in $SU(2)$ is $2\alpha_1$. Therefore, since  $\tr(U_1^\dagger U_2)=0$ for $U_1,U_2$ orthogonal, and $U_1^\dagger U_2$ is in the set of unitary 2x2 matrices, this implies that if $U_1$ maps to ${\cal O}_1$ and $U_2$ maps to ${\cal O}_2$, then orthogonal $U_1,U_2$ implies that $\tr({\cal O}^T_1{\cal O}_2)=-1$.  This is the equivalent orthogonality condition for the $SO(3)$ matrices.  This is useful for affine maps of the polarization vector, as seen, for example, in  (\cite{Nielsen_Chuang_Textbook_2011,Byrd_2011}).


\section{What if we had access to only one qubit}
\label{Sec:One_Qubit}

With only having access to one qubit of a two-qubit system, we can apply either a left or a right special orthogonal rotation $\mathcal{O} \in $ SO(3) on the correlation matrix $\mathcal{T}_\rho$ which prevents us from always being able to diagonalize it. On the other hand, the QR decomposition (see Chapter 2 of \cite{horn_johnson_2013} for a general discussion) allows us to write 
\begin{align}\label{eq:qr_decomp}
    \mathcal{T}_\rho=QR,
\end{align}
where $Q$ is orthonormal and $R$ is upper triangular.
The QR decomposition for the correlation matrix Eq.~\eqref{eq:qr_decomp} allows us to perform the Gram-Schmidt process. We can apply one orthogonal rotation to put it into upper or lower triangular form. The orthogonal rotation that needs to be applied on the left side will have the form $Q^T$ and the orthogonal rotation that needs to be applied on the right side will have the form $Q$. Thus, for a given correlation matrix $\mathcal{T}_\rho$ we can determine $Q$ and use the results of Eq. \eqref{Eq:General_Solution} to determine the local unitary rotation to rotate the correlation matrix to upper or lower triangular.

We can also design the correlation matrix to be symmetric from only having access to one of the qubits. For instance, if the correlation matrix $\mathcal{T}_\rho$ has the form $L \Sigma R \in SO(3) \otimes D \otimes SO(3)$ for a diagonal matrix $\Sigma$, then we would simply want to induce $R^T L^T$ using local unitaries on either one of the systems so that we obtain $R^T \Sigma R$ or $L \Sigma L^T$ respectively. Symmetric correlation matrices have the property that measurements of expectation values of local observables $\omega\otimes\sigma$ (local to the first (second) system) are identical to the
expectation values of $\sigma\otimes\omega$. Thus, an outside observer cannot distinguish which of these two scenarios was performed.

An intermediate resource that falls between entanglement and Bell nonlocality is called quantum steering \cite{Jevtic_2014,Bowles_2014, Nguyen_2020QuantumSteeringBellDiagStatesGenMeas, Sun_2017, Gheorghiu_2017RigidityOfQuantSteeringAndOneSidedDeviceIndVerQC}. Similar to a local hidden variable model for Bell inequalities, there may exist a local hidden state model that can describe Bob's marginal distribution after Alice has performed measurements on a distant qubit that is entangled to Bob's qubit. States that have a local hidden state model that describes Bob's marginal distribution after Alice's measurement are unsteerable. Quantum steering inequalities can be used to detect entanglement \cite{Sun_2017}.  


\section{Data Availability}\label{sec:DataAvail}
The python and Mathematica codes used to analyze our explicit formula during the current study are available in the SO-3-to-SU2- repository on github, \href{https://github.com/quantumdilley/SO-3-to-SU-2-.git}{https://github.com/quantumdilley/SO-3-to-SU-2-.git}.

\section{Conclusion}\label{sec:conclusion}

In this paper, we have given an explicit mapping that takes any element of SO(3) as its input and gives the associated elements of SU(2).  This gives the form of a unitary transformation on a two-qubit transformation that would be required to produce a given SO(3) operator. This allows us to determine the correct local unitaries that diagonalize the correlation matrix of a two-level quantum state without having to solve for the parameters in SO(3) explicitly. There is already a well-known exact solution for the reverse direction, but we provide a closed-form solution that gives us the capability of guiding the state when given an SO(3) action. This transformation is more complex since there is a double cover of SO(3) by SU(2) and also due to the isomorphism between SU(2) and unit quaternions. When the unit quaternions contained a real part, the solution was simple and given explicitly by the $U_R(\mathcal{O})$ part in Eq. (\ref{Eq:General_Solution}). When the unit quaternions turn solely into vector quaternions, we needed to solve for each individual case directly as we have shown in Appendix \ref{App:V_Sol_Proof}.

We were able to determine the orthogonality conditions on both the local special unitary operators and the corresponding special orthogonal matrices that can be useful for the affine maps of the polarization vector \cite{Byrd_2011}. Furthermore, we discussed the implications when access to only one qubit of a correlated two-qubit state is available. In this case, the correlation matrix cannot be diagonalized. Albeit the circumstance, if we have access to any SO(3) rotation on our local system, then we have the ability to make the correlation matrix symmetric with knowledge of the other local system. We can also perform the QR-decomposition to make the correlation matrix upper or lower triangular; depending on which system is controllable. There are many instances where one lab has only partial access to a quantum state. This work provides the details of how that control can be accomplished.

\section{Acknowledgments}\label{sec:acknowledgements}
Funding for this research was provided by the NSF, MPS under award number PHYS-1820870. 

\appendix

\section{Appendix}\label{sec:appendix}

\subsection{Correlation matrices for the Bell states.}
To determine the correlation matrix of any two-qubit density operator, simply perform the calculations $\tr(\hat{\sigma}_i \otimes \hat{\sigma_j} \cdot \rho) = \{\mathcal{T}_{\rho}\}_{ij}$. Using this formula, we can directly determine the correlation matrices for the maximally entangled Bell operators:
\begin{align}\label{eq:corr_mats1}
    \mathcal{T}_{\Phi^+} &=\
    \left(
    \begin{array}{ccc}
        1 & 0 & 0 \\
        0 & -1 & 0 \\
        0 & 0 & 1
    \end{array}
    \right) 
    \quad
    \mathcal{T}_{\Phi^-} =\
    \left(
    \begin{array}{ccc}
        -1 & 0 & 0 \\
        0 & 1 & 0 \\
        0 & 0 & 1
    \end{array}
    \right) \\
    \label{eq:corr_mats2}
    \mathcal{T}_{\Psi^+} &=\
    \left(
    \begin{array}{ccc}
        1 & 0 & 0 \\
        0 & 1 & 0 \\
        0 & 0 & -1
    \end{array}
    \right) 
    \quad
    \mathcal{T}_{\Psi^-} =\
    \left(
    \begin{array}{ccc}
        -1 & 0 & 0 \\
        0 & -1 & 0 \\
        0 & 0 & -1
    \end{array}
    \right).
\end{align}

\subsection{Proof of equation \eqref{Eq:V_Sol}}
\label{App:V_Sol_Proof}
Let us now calculate each part of Eq. (\ref{Eq:V_Sol}) case by case. Note that for all these cases $\alpha_1=0$ as seen from Eq. \eqref{eq:tracecond}.
\newline
\begin{center}
    \textbf{Case 1: }  $\text{sgn}(\alpha_2)\text{sgn}(\beta_1)\text{sgn}(\beta_2) \neq 0$
\end{center}
\begin{align}
    W(\mathcal{O},\abs{L_1},\abs{L_2},\abs{L_3}) =
    \eta_1\left(
    \begin{array}{cc}
        i \alpha_2  & \beta_1+i \beta_2 \\
        -\beta_1 + i \beta_2 & -i \alpha_2
    \end{array}
    \right)
\end{align}
where $\eta_1 = \text{sgn}(\alpha_2)\text{sgn}(\beta_1)\text{sgn}(\beta_2)$.
\newline
\begin{center}
    \textbf{Case 2: }  $\text{sgn}(\alpha_2) = 0 \; \text{and} \; \text{sgn}(\beta_1)\text{sgn}(\beta_2) \neq 0$
\end{center}
\begin{align}
    W(\mathcal{O},\mathbb{I},\mathbb{I},-\abs{L_1}) =
    \eta_2\left(
    \begin{array}{cc}
        -i\abs{\alpha_2}\eta_2  & \beta_1 + i \beta_2 \\
        -\beta_1 + i \beta_2 & i\abs{\alpha_2}\eta_2
    \end{array}
    \right)
\end{align}
where $\eta_2 = -\text{sgn}(\beta_1)$. Since $\text{sgn}(\alpha_2) = 0$ implies that $\alpha_2 = 0$, this form is correct. The $\eta_2$ only adds a $\pm$ global phase. We also see that
\begin{align}
    (1-\gamma_1)\gamma_2\gamma_3 = (1-\text{sgn}(\alpha_2 \beta_1)^2)(1-\text{sgn}(\alpha_2 \beta_2)^2)\text{sgn}(\beta_1 \beta_2)^2
\end{align}
which can be expressed in cases as
\begin{align}
   (1-\gamma_1)\gamma_2\gamma_3 =
    \begin{cases}
        1 &\text{ if } \text{sgn}(\alpha_2) = 0 \; \text{and} \; \text{sgn}(\beta_1)\text{sgn}(\beta_2) \neq 0 \\
        0 &\text{ otherwise}
    \end{cases}.
\end{align}
\begin{center}
    \textbf{Case 3: }  $\text{sgn}(\beta_1) = 0 \; \& \; \text{sgn}(\alpha_2)\text{sgn}(\beta_2) \neq 0$
\end{center}
\begin{align}
    W(\mathcal{O},-\abs{L_2},\mathbb{I},\mathbb{I}) =
    \eta_3\left(
    \begin{array}{cc}
        i\alpha_2 & -\abs{\beta_1} \eta_3 + i \beta_2 \\
        \abs{\beta_1} \eta_3 + i \beta_2 & -i\alpha_2
    \end{array}
    \right)
\end{align}
where $\eta_3 = -\text{sgn}(\beta_2)$. Since $\text{sgn}(\beta_1) = 0$ implies that $\beta_1 = 0$, this form is correct. The $\eta_3$ only adds a $\pm$ global phase. We also see that
\begin{align}
    \gamma_1(1-\gamma_2)\gamma_3 = (1-\text{sgn}(\alpha_2 \beta_1)^2)(1-\text{sgn}(\beta_1 \beta_2)^2)\text{sgn}(\alpha_2 \beta_2)^2
\end{align}
which can be expressed in cases as
\begin{align}
   \gamma_1(1-\gamma_2)\gamma_3 =
    \begin{cases}
        1 &\text{ if } \text{sgn}(\beta_1) = 0 \; \text{and} \; \text{sgn}(\alpha_2)\text{sgn}(\beta_2) \neq 0 \\
        0 &\text{ otherwise}
    \end{cases}.
\end{align}
\begin{center}
    \textbf{Case 4: }  $\text{sgn}(\beta_2) = 0 \; \& \; \text{sgn}(\alpha_2)\text{sgn}(\beta_1) \neq 0$
\end{center}
\begin{align}
    W(\mathcal{O},\mathbb{I},-\abs{L_3},\mathbb{I}) =
    \eta_4\left(
    \begin{array}{cc}
        i\alpha_2 & \beta_1 - i \abs{\beta_2} \eta_4 \\
        -\beta_1 - i \abs{\beta_2} \eta_4 & -i\alpha_2
    \end{array}
    \right)
\end{align}
where $\eta_4 = -\text{sgn}(\alpha_2)$. Since $\text{sgn}(\beta_2) = 0$ implies that $\beta_2 = 0$, this form is correct. The $\eta_4$ only adds a $\pm$ global phase. We also see that
\begin{align}
    \gamma_1 \gamma_2 (1-\gamma_3) = (1-\text{sgn}(\alpha_2 \beta_2)^2)(1-\text{sgn}(\beta_1 \beta_2)^2)\text{sgn}(\alpha_2 \beta_1)^2
\end{align}
which can be expressed in cases as
\begin{align}
    \gamma_1 \gamma_2 (1-\gamma_3) =
    \begin{cases}
        1 &\text{ if } \text{sgn}(\beta_2) = 0 \; \& \; \text{sgn}(\alpha_2)\text{sgn}(\beta_1) \neq 0 \\
        0 &\text{ otherwise}
    \end{cases}.
\end{align}
\begin{align} \nonumber
     \textbf{Case 5: }  \text{sgn}(\alpha_2) \neq 0 \; \& \; \text{sgn}(\beta_1),\text{sgn}(\beta_2) = 0 \\ \nonumber
     \textbf{Case 6: }  \text{sgn}(\beta_1) \neq 0 \; \& \; \text{sgn}(\alpha_2),\text{sgn}(\beta_2) = 0 \\
     \textbf{Case 7: }  \text{sgn}(\beta_2) \neq 0 \; \& \; \text{sgn}(\alpha_2),\text{sgn}(\beta_1) = 0
\end{align}
\begin{align}
    W(\mathcal{O},-\mathbb{I},-\mathbb{I},-\mathbb{I}) =
    \left(
    \begin{array}{cc}
        i\abs{\alpha_2} & \abs{\beta_1} + i \abs{\beta_2} \\
        -\abs{\beta_1} + i \abs{\beta_2} & -i\abs{\alpha_2}
    \end{array}
    \right).
\end{align}
Since only one of the elements of $\{\alpha_2, \beta_1, \beta_2\}$ are nonzero, this form is correct. The missing sign is only a $\pm$ global phase. We also see that
\begin{align}
    \gamma_1\gamma_2\gamma_3 = (1-\text{sgn}(\alpha_2 \beta_1)^2)(1-\text{sgn}(\alpha_2 \beta_2)^2)(1-\text{sgn}(\beta_1 \beta_2)^2)
\end{align}
which can be expressed in cases as
\begin{align}
   \gamma_1\gamma_2\gamma_3 =
    \begin{cases}
        1 &\text{ if } \text{sgn}(\alpha_2),\text{sgn}(\beta_1),\text{sgn}(\beta_2) = 0 \\
        &\text{ if } \text{sgn}(\alpha_2),\text{sgn}(\beta_1) = 0 \; \& \; \text{sgn}(\beta_2) \neq 0 \\
        &\text{ if } \text{sgn}(\alpha_2),\text{sgn}(\beta_2) = 0 \; \& \; \text{sgn}(\beta_1) \neq 0 \\
        &\text{ if } \text{sgn}(\beta_1),\text{sgn}(\beta_2) = 0 \; \& \; \text{sgn}(\alpha_2) \neq 0 \\
        0 &\text{ otherwise}
    \end{cases},
\end{align}
which completes the rest of the cases involved when $\tr(\mathcal{O}) = -1$. The $\gamma$ functions ensure that there are no repeats of any solutions in Eq. (\ref{Eq:V_Sol}). Now we can safely say that all of the 8 possible cases described by Eq. (\ref{Eq:General_Solution}) have been proven. Case 8 is when $\tr(\mathcal{O}) \neq -1$ and it has been proven in Eq. (\ref{Eq:U_Sol_General}).




\begin{thebibliography}{26}
\ifx \bisbn   \undefined \def \bisbn  #1{ISBN #1}\fi
\ifx \binits  \undefined \def \binits#1{#1}\fi
\ifx \bauthor  \undefined \def \bauthor#1{#1}\fi
\ifx \batitle  \undefined \def \batitle#1{#1}\fi
\ifx \bjtitle  \undefined \def \bjtitle#1{#1}\fi
\ifx \bvolume  \undefined \def \bvolume#1{\textbf{#1}}\fi
\ifx \byear  \undefined \def \byear#1{#1}\fi
\ifx \bissue  \undefined \def \bissue#1{#1}\fi
\ifx \bfpage  \undefined \def \bfpage#1{#1}\fi
\ifx \blpage  \undefined \def \blpage #1{#1}\fi
\ifx \burl  \undefined \def \burl#1{\textsf{#1}}\fi
\ifx \doiurl  \undefined \def \doiurl#1{\url{https://doi.org/#1}}\fi
\ifx \betal  \undefined \def \betal{\textit{et al.}}\fi
\ifx \binstitute  \undefined \def \binstitute#1{#1}\fi
\ifx \binstitutionaled  \undefined \def \binstitutionaled#1{#1}\fi
\ifx \bctitle  \undefined \def \bctitle#1{#1}\fi
\ifx \beditor  \undefined \def \beditor#1{#1}\fi
\ifx \bpublisher  \undefined \def \bpublisher#1{#1}\fi
\ifx \bbtitle  \undefined \def \bbtitle#1{#1}\fi
\ifx \bedition  \undefined \def \bedition#1{#1}\fi
\ifx \bseriesno  \undefined \def \bseriesno#1{#1}\fi
\ifx \blocation  \undefined \def \blocation#1{#1}\fi
\ifx \bsertitle  \undefined \def \bsertitle#1{#1}\fi
\ifx \bsnm \undefined \def \bsnm#1{#1}\fi
\ifx \bsuffix \undefined \def \bsuffix#1{#1}\fi
\ifx \bparticle \undefined \def \bparticle#1{#1}\fi
\ifx \barticle \undefined \def \barticle#1{#1}\fi
\bibcommenthead
\ifx \bconfdate \undefined \def \bconfdate #1{#1}\fi
\ifx \botherref \undefined \def \botherref #1{#1}\fi
\ifx \url \undefined \def \url#1{\textsf{#1}}\fi
\ifx \bchapter \undefined \def \bchapter#1{#1}\fi
\ifx \bbook \undefined \def \bbook#1{#1}\fi
\ifx \bcomment \undefined \def \bcomment#1{#1}\fi
\ifx \oauthor \undefined \def \oauthor#1{#1}\fi
\ifx \citeauthoryear \undefined \def \citeauthoryear#1{#1}\fi
\ifx \endbibitem  \undefined \def \endbibitem {}\fi
\ifx \bconflocation  \undefined \def \bconflocation#1{#1}\fi
\ifx \arxivurl  \undefined \def \arxivurl#1{\textsf{#1}}\fi
\csname PreBibitemsHook\endcsname

\bibitem{Bennet_1993}
\begin{barticle}
\bauthor{\bsnm{Bennett}, \binits{C.H.}},
\bauthor{\bsnm{Brassard}, \binits{G.}},
\bauthor{\bsnm{Cr\'epeau}, \binits{C.}},
\bauthor{\bsnm{Jozsa}, \binits{R.}},
\bauthor{\bsnm{Peres}, \binits{A.}},
\bauthor{\bsnm{Wootters}, \binits{W.K.}}:
\batitle{Teleporting an unknown quantum state via dual classical and
  einstein-podolsky-rosen channels}.
\bjtitle{Phys. Rev. Lett.}
\bvolume{70},
\bfpage{1895}--\blpage{1899}
(\byear{1993}).
\doiurl{10.1103/PhysRevLett.70.1895}
\end{barticle}
\endbibitem

\bibitem{Bennet_1992}
\begin{barticle}
\bauthor{\bsnm{Bennett}, \binits{C.H.}},
\bauthor{\bsnm{Wiesner}, \binits{S.J.}}:
\batitle{Communication via one- and two-particle operators on
  einstein-podolsky-rosen states}.
\bjtitle{Phys. Rev. Lett.}
\bvolume{69},
\bfpage{2881}--\blpage{2884}
(\byear{1992}).
\doiurl{10.1103/PhysRevLett.69.2881}
\end{barticle}
\endbibitem

\bibitem{Grover_1996}
\begin{bchapter}
\bauthor{\bsnm{Grover}, \binits{L.K.}}:
\bctitle{A fast quantum mechanical algorithm for database search}.
In: \bbtitle{Proceedings of the Twenty-Eighth Annual ACM Symposium on Theory of
  Computing}.
\bsertitle{STOC '96},
pp. \bfpage{212}--\blpage{219}.
\bpublisher{Association for Computing Machinery},
\blocation{New York, NY, USA}
(\byear{1996}).
\doiurl{10.1145/237814.237866}.
\burl{https://doi.org/10.1145/237814.237866}
\end{bchapter}
\endbibitem

\bibitem{Shor_1994}
\begin{bchapter}
\bauthor{\bsnm{Shor}, \binits{P.W.}}:
\bctitle{Algorithms for quantum computation: discrete logarithms and
  factoring}.
In: \bbtitle{Proceedings 35th Annual Symposium on Foundations of Computer
  Science},
pp. \bfpage{124}--\blpage{134}
(\byear{1994}).
\doiurl{10.1109/SFCS.1994.365700}
\end{bchapter}
\endbibitem

\bibitem{Parakh_2022}
\begin{barticle}
\bauthor{\bsnm{{Parakh}}, \binits{A.}}:
\batitle{{Quantum teleportation with one classical bit}}.
\bjtitle{Scientific Reports}
\bvolume{12},
\bfpage{3392}
(\byear{2022})
{\href{https://arxiv.org/abs/2110.11254}{{arXiv:2110.11254}}}
{[quant-ph]}.
\doiurl{10.1038/s41598-022-06853-w}
\end{barticle}
\endbibitem

\bibitem{cerf_2000}
\begin{barticle}
\bauthor{\bsnm{Cerf}, \binits{N.J.}},
\bauthor{\bsnm{Gisin}, \binits{N.}},
\bauthor{\bsnm{Massar}, \binits{S.}}:
\batitle{Classical teleportation of a quantum bit}.
\bjtitle{Phys. Rev. Lett.}
\bvolume{84},
\bfpage{2521}--\blpage{2524}
(\byear{2000}).
\doiurl{10.1103/PhysRevLett.84.2521}
\end{barticle}
\endbibitem

\bibitem{Gottesman_1999}
\begin{barticle}
\bauthor{\bsnm{{Gottesman}}, \binits{D.}},
\bauthor{\bsnm{{Chuang}}, \binits{I.L.}}:
\batitle{{Demonstrating the viability of universal quantum computation using
  teleportation and single-qubit operations}}.
\bjtitle{Nature}
\bvolume{402}(\bissue{6760}),
\bfpage{390}--\blpage{393}
(\byear{1999})
{[quant-ph]}.
\doiurl{10.1038/46503}
\end{barticle}
\endbibitem

\bibitem{Blume_Kohout_2010}
\begin{barticle}
\bauthor{\bsnm{Blume-Kohout}, \binits{R.}}:
\batitle{Optimal, reliable estimation of quantum states}.
\bjtitle{New Journal of Physics}
\bvolume{12}(\bissue{4}),
\bfpage{043034}
(\byear{2010}).
\doiurl{10.1088/1367-2630/12/4/043034}
\end{barticle}
\endbibitem

\bibitem{Li_2017}
\begin{barticle}
\bauthor{\bsnm{Li}, \binits{M.}},
\bauthor{\bsnm{Xue}, \binits{G.}},
\bauthor{\bsnm{Tan}, \binits{X.}},
\bauthor{\bsnm{Liu}, \binits{Q.}},
\bauthor{\bsnm{Dai}, \binits{K.}},
\bauthor{\bsnm{Zhang}, \binits{K.}},
\bauthor{\bsnm{Yu}, \binits{H.}},
\bauthor{\bsnm{Yu}, \binits{Y.}}:
\batitle{Two-qubit state tomography with ensemble average in coupled
  superconducting qubits}.
\bjtitle{Applied Physics Letters}
\bvolume{110}(\bissue{13}),
\bfpage{132602}
(\byear{2017}).
\doiurl{10.1063/1.4979652}
\end{barticle}
\endbibitem

\bibitem{Horodecki_1995}
\begin{barticle}
\bauthor{\bsnm{Horodecki}, \binits{R.}},
\bauthor{\bsnm{Horodecki}, \binits{P.}},
\bauthor{\bsnm{Horodecki}, \binits{M.}}:
\batitle{Violating bell inequality by mixed spin-12 states: necessary and
  sufficient condition}.
\bjtitle{Physics Letters A}
\bvolume{200}(\bissue{5}),
\bfpage{340}--\blpage{344}
(\byear{1995}).
\doiurl{10.1016/0375-9601(95)00214-N}
\end{barticle}
\endbibitem

\bibitem{Hyllus_2005}
\begin{barticle}
\bauthor{\bsnm{Hyllus}, \binits{P.}},
\bauthor{\bsnm{G\"uhne}, \binits{O.}},
\bauthor{\bsnm{Bru\ss{}}, \binits{D.}},
\bauthor{\bsnm{Lewenstein}, \binits{M.}}:
\batitle{Relations between entanglement witnesses and bell inequalities}.
\bjtitle{Phys. Rev. A}
\bvolume{72},
\bfpage{012321}
(\byear{2005}).
\doiurl{10.1103/PhysRevA.72.012321}
\end{barticle}
\endbibitem

\bibitem{Zhang_2013}
\begin{barticle}
\bauthor{\bsnm{Zhang}, \binits{T.-M.}},
\bauthor{\bsnm{Wu}, \binits{R.-B.}}:
\batitle{Minimum-time control of local quantum gates for two-qubit homonuclear
  systems}.
\bjtitle{IFAC Proceedings Volumes}
\bvolume{46}(\bissue{20}),
\bfpage{359}--\blpage{364}
(\byear{2013}).
\doiurl{10.3182/20130902-3-CN-3020.00031}.
\bcomment{3rd IFAC Conference on Intelligent Control and Automation Science
  ICONS 2013}
\end{barticle}
\endbibitem

\bibitem{Khaneja_2001}
\begin{barticle}
\bauthor{\bsnm{Khaneja}, \binits{N.}},
\bauthor{\bsnm{Brockett}, \binits{R.}},
\bauthor{\bsnm{Glaser}, \binits{S.J.}}:
\batitle{Time optimal control in spin systems}.
\bjtitle{Phys. Rev. A}
\bvolume{63},
\bfpage{032308}
(\byear{2001}).
\doiurl{10.1103/PhysRevA.63.032308}
\end{barticle}
\endbibitem

\bibitem{cornwell1984}
\begin{bbook}
\bauthor{\bsnm{Cornwell}, \binits{J.F.}}:
\bbtitle{Group Theory in Physics}.
\bsertitle{Group Theory in Physics},
vol. \bseriesno{v. 2}.
\bpublisher{Academic Press},
\blocation{Cambridge, Massachusetts}
(\byear{1984}).
\burl{https://books.google.com/books?id=bKQ7AQAAIAAJ}
\end{bbook}
\endbibitem

\bibitem{Makhlin_2002}
\begin{barticle}
\bauthor{\bsnm{{Makhlin}}, \binits{Y.}}:
\batitle{Nonlocal properties of two-qubit gates and mixed states and
  optimization of quantum computations}.
\bjtitle{Quantum Information Processing}
\bvolume{1},
\bfpage{243}--\blpage{252}
(\byear{2002}).
\doiurl{10.1023/A:1022144002391}
\end{barticle}
\endbibitem

\bibitem{Hamilton_1843}
\begin{botherref}
\oauthor{\bsnm{Hamilton}, \binits{R.}}:
On quaternions; or on a new system of imaginaries in algebra
(1843)
\end{botherref}
\endbibitem

\bibitem{Euler_1771}
\begin{barticle}
\bauthor{\bsnm{Euler}, \binits{L.}}:
\batitle{Problema algebraicum ob affectiones prorsus singulares memorabile}.
\bjtitle{Commentatio 407 indicis Enestrœmiani, Novi commentarii academiæ
  scientiarum Petropolitanæ}
\bvolume{15}(\bissue{407}),
\bfpage{75}--\blpage{106}
(\byear{1771})
\end{barticle}
\endbibitem

\bibitem{Hall_2015}
\begin{bbook}
\bauthor{\bsnm{Hall}, \binits{B.}}:
\bbtitle{Lie Groups, Lie Algebras, and Representations: An Elementary
  Introduction, Graduate Texts in Mathematics},
\bedition{2}nd edn.
\bpublisher{Springer},
\blocation{New York, NY, USA}
(\byear{2015})
\end{bbook}
\endbibitem

\bibitem{Byrd_2011}
\begin{barticle}
\bauthor{\bsnm{Byrd}, \binits{M.S.}},
\bauthor{\bsnm{Bishop}, \binits{C.A.}},
\bauthor{\bsnm{Ou}, \binits{Y.-C.}}:
\batitle{General open-system quantum evolution in terms of affine maps of the
  polarization vector}.
\bjtitle{Phys. Rev. A}
\bvolume{83},
\bfpage{012301}
(\byear{2011}).
\doiurl{10.1103/PhysRevA.83.012301}
\end{barticle}
\endbibitem

\bibitem{Nielsen_Chuang_Textbook_2011}
\begin{bbook}
\bauthor{\bsnm{Nielsen}, \binits{M.A.}},
\bauthor{\bsnm{Chuang}, \binits{I.L.}}:
\bbtitle{Quantum Computation and Quantum Information: 10th Anniversary
  Edition},
\bedition{10th} edn.
\bpublisher{Cambridge University Press},
\blocation{New York, NY, USA}
(\byear{2011})
\end{bbook}
\endbibitem

\bibitem{horn_johnson_2013}
\begin{bbook}
\bauthor{\bsnm{Horn}, \binits{R.A.}},
\bauthor{\bsnm{Johnson}, \binits{C.R.}}:
\bbtitle{Matrix Analysis}.
\bpublisher{Cambridge University Press},
\blocation{New York, NY}
(\byear{2013})
\end{bbook}
\endbibitem

\bibitem{Jevtic_2014}
\begin{barticle}
\bauthor{\bsnm{Jevtic}, \binits{S.}},
\bauthor{\bsnm{Pusey}, \binits{M.}},
\bauthor{\bsnm{Jennings}, \binits{D.}},
\bauthor{\bsnm{Rudolph}, \binits{T.}}:
\batitle{Quantum steering ellipsoids}.
\bjtitle{Phys. Rev. Lett.}
\bvolume{113},
\bfpage{020402}
(\byear{2014}).
\doiurl{10.1103/PhysRevLett.113.020402}
\end{barticle}
\endbibitem

\bibitem{Bowles_2014}
\begin{barticle}
\bauthor{\bsnm{Bowles}, \binits{J.}},
\bauthor{\bsnm{V\'ertesi}, \binits{T.}},
\bauthor{\bsnm{Quintino}, \binits{M.T.}},
\bauthor{\bsnm{Brunner}, \binits{N.}}:
\batitle{One-way einstein-podolsky-rosen steering}.
\bjtitle{Phys. Rev. Lett.}
\bvolume{112},
\bfpage{200402}
(\byear{2014}).
\doiurl{10.1103/PhysRevLett.112.200402}
\end{barticle}
\endbibitem

\bibitem{Nguyen_2020QuantumSteeringBellDiagStatesGenMeas}
\begin{barticle}
\bauthor{\bsnm{Nguyen}, \binits{H.C.}},
\bauthor{\bsnm{G\"uhne}, \binits{O.}}:
\batitle{Quantum steering of bell-diagonal states with generalized
  measurements}.
\bjtitle{Phys. Rev. A}
\bvolume{101},
\bfpage{042125}
(\byear{2020}).
\doiurl{10.1103/PhysRevA.101.042125}
\end{barticle}
\endbibitem

\bibitem{Sun_2017}
\begin{barticle}
\bauthor{\bsnm{{Sun}}, \binits{W.-Y.}},
\bauthor{\bsnm{{Wang}}, \binits{D.}},
\bauthor{\bsnm{{Shi}}, \binits{J.-D.}},
\bauthor{\bsnm{{Ye}}, \binits{L.}}:
\batitle{{Exploration quantum steering, nonlocality and entanglement of
  two-qubit X-state in structured reservoirs}}.
\bjtitle{Scientific Reports}
\bvolume{7},
\bfpage{39651}
(\byear{2017})
{[quant-ph]}.
\doiurl{10.1038/srep39651}
\end{barticle}
\endbibitem

\bibitem{Gheorghiu_2017RigidityOfQuantSteeringAndOneSidedDeviceIndVerQC}
\begin{barticle}
\bauthor{\bsnm{Gheorghiu}, \binits{A.}},
\bauthor{\bsnm{Wallden}, \binits{P.}},
\bauthor{\bsnm{Kashefi}, \binits{E.}}:
\batitle{Rigidity of quantum steering and one-sided device-independent
  verifiable quantum computation}.
\bjtitle{New Journal of Physics}
\bvolume{19}(\bissue{2}),
\bfpage{023043}
(\byear{2017}).
\doiurl{10.1088/1367-2630/aa5cff}
\end{barticle}
\endbibitem

\end{thebibliography}
\end{document}